\documentclass[12pt]{article}
\usepackage{graphics}
\usepackage{amssymb}
\usepackage{amsmath}

\newcommand{\rf}[1]{(\ref{#1})}
\newcommand{\beq}{\begin{equation}}
\newcommand{\eeq}{\end{equation}}
\newcommand{\bea}{\begin{eqnarray}}
\newcommand{\eea}{\end{eqnarray}}

\newcommand{\e}{\mbox{e}}
\renewcommand{\d}{\mbox{d}}
\newcommand{\g}{\gamma}

\renewcommand{\a}{\alpha}

\newcommand{\m}{\mu}

\newcommand{\ra}{\rangle}
\newcommand{\la}{\langle}

\newcommand{\cO}{{\cal O}}

\newcommand{\tZ}{{\tilde{Z}}}

\begin{document}

\begin{center}
\vspace{24pt}
{ \large \bf Baby Universes Revisited}

\vspace{30pt}

{\sl J. Ambj\o rn}$\,^{a}$,
{\sl J. Barkley}$\,^{a}$,
{\sl T. Budd}$\,^{b}$
and {\sl R. Loll}$\,^{b}$

\vspace{48pt}
{\footnotesize

$^a$~The Niels Bohr Institute, Copenhagen University\\
Blegdamsvej 17, DK-2100 Copenhagen \O , Denmark.\\
email: ambjorn@nbi.dk, barkley@nbi.dk\\

\vspace{10pt}

$^b$~Institute for Theoretical Physics, Utrecht University, \\
Leuvenlaan 4, NL-3584 CE Utrecht, The Netherlands.\\
email: t.g.budd@uu.nl, r.loll@uu.nl\\

}
\vspace{96pt}
\end{center}


\begin{center}
{\bf Abstract}
\end{center}

\noindent The behaviour of baby universes has been an important ingredient in understanding
and quantifying non-critical string theory or, equivalently, models of two-dimensional
Euclidean quantum gravity coupled to matter. Within a regularized description based
on dynamical triangulations, we amend an earlier conjecture by Jain and Mathur on
the scaling behaviour of genus-$g$ surfaces containing particular baby universe `necks', and
perform a nontrivial numerical check on our improved conjecture.

\vspace{12pt}
\noindent

\vspace{24pt}
\noindent
PACS: 04.60.Ds, 04.60.Kz, 04.06.Nc, 04.62.+v.\\
Keywords: quantum gravity, lower dimensional models, lattice models.

\newpage

\section{Introduction}\label{intro}

In physics, the intriguing concept of baby universes as the offspring of a parent
universe -- acquiring a life of their own while remaining connected to the 
latter -- has been most concretely 
realized in 2d Euclidean quantum gravity models. 
In these models
the partition function for a compact genus-$g$ surface coupled to conformal 
matter with central charge $c$ is given by
\beq\label{1a}
Z^{(g)}(A) = A^{\g(g)-3} \e^{\m A},
\eeq
a formula which is exact at the conformal point (see e.g. \cite{dk}).
In (\ref{1a}), $A$ denotes the area of the surface, $\m$ a bare 
(cutoff-dependent) cosmological constant, and $\g(g)$
the so-called susceptibility exponent or entropy exponent, given by
\beq\label{2a}
\g(g) = \g_0 +g(2-\g_0),~~~~\g_0=\g(0)= \frac{c-1 -\sqrt{(c-1)(c-25)}}{12}.
\eeq 
For $c=0$, where no matter is present, 
the partition function \rf{1a} has an entropic interpretation, namely,
it ``counts" the number of surfaces of genus $g$ with area $A$. After  
introducing a regularization of the continuum theory, for example,
by using dynamical triangulations (DT), this counting becomes literal. 
The continuum area
is then $A = \sqrt{3} a^2 N/4$, where $N$ is the number of triangles,
and $a$ the edge length of the equilateral triangles used 
in DT, simultaneously acting as a UV-cutoff of the theory. 
The number of distinct piecewise linear surfaces (triangulations) of genus $g$
one can construct from $N$ such triangles is
\beq\label{3a}
Z^{(g)}(N) = k(g) \, N^{\g(g)-3} \e^{\mu_0 N}\left(1+\cO(1/N)\right),
\eeq  
where $\m_0$ is independent of the genus $g$ and $k(g)$ is a constant.

In an influential paper, Jain and Mathur (JM) related the entropy
exponent $\g_0$ to the creation of baby universes \cite{jm}.
They showed that the average number of so-called ``minbus'' 
(minimal neck baby universes of disc topology) of regularized area $B$ 
is given by
\beq\label{4a}
n_A(B) = k(g)\, A \left(1-B/A\right)^{\g(g)-2} \, B^{\g_0-2},
\eeq
where $A$ is the regularized area of the total surface, and we assume 
$B<A/2$. Cutting the surface 
open along the minimal neck will produce two disconnected
surfaces, a genus-$g$ surface of area $A\!-\! B$ with 
one $S^1$-boundary (the minimal neck), and a baby universe of area $B$, 
whose topology is a sphere with a single
$S^1$-boundary 
(the same minimal neck), that is, a disc. 
In the following we will always work in a suitable 
DT-ensemble of triangulated surfaces and use the notation 
$A$ and $B$ for
the number of triangles, thus ignoring the factor $a^2\sqrt{3}/4$ which 
relates the number of triangles to the area of the surface. 
Likewise in discrete units, the length of a minimal neck can be 
1, 2 or 3 links ($\equiv$ triangle edges), de\-pen\-ding on the choice of DT-ensemble,
i.e. the regularity conditions imposed on the triangle gluings.

The leading power-law behaviour of $B$ in \rf{4a} gives us a convenient 
way to determine $\g_0$
in numerical simulations and has been used extensively in 
two dimensions\footnote{``Integrating out'' baby 
universes in two dimensions \cite{ackl} leads to an alternative 
theory of 2d quantum gravity
known as ``causal dynamical triangulations'' (CDT) \cite{al}. By
introducing a coupling constant for baby universe
creation one can relate the two models \cite{mod-cdt}.}, 
beginning with \cite{ajt}, as well as in higher-dimensional gravity, 
starting with \cite{ajjk}.
  
While the derivation of \rf{4a} is robust and correct, 
JM put forward generalized relations, based on a certain conjectured formula   
involving non-minimal necks (eq.\ (9) in \cite{jm}), which also divide a surface 
into two pieces
$A-B$ and $B$, but have an arbitrary length $\ell < \sqrt{A}$. 
The main purpose of this letter is to demonstrate that the formula conjectured 
in \cite{jm} is incorrect. We will discuss how to 
modify it appropriately, while retaining many of the results derived there. 
Our key observation
will be that the necks of baby universes are by construction rather special curves
on the two-surface, whose scaling behaviour (as function of the surface
area $A$) is different from that of generic curves. 
We will also perform a nontrivial numerical check of our improved conjecture and
suggest further applications of our new result.

\section{The conjecture and how to modify it}

In this section we will limit ourselves to the case $c\! =\! 0$ of 
2d gravity without matter, leaving the discussion of general 
central charge $c\leq 1$ to Sec.\ \ref{sec3}.  

Consider a (triangulated) surface of area $A$ and spherical
topology, except for $n$ boundary loops of length $\ell_i$, $i=1,\ldots,n$,
counting the number of edges in the boundaries.
Assume also that none of the boundary loops can be deformed into  
a loop of shorter length in the same homotopy class, 
unless the deformation sweeps an area 
which is a sizeable fraction of $A$. JM conjectured
that the number of such surfaces behaves like 
\beq\label{1b}
\tZ_n(A,\ell_i)\sim Z^{(g=0)}(A)\; A^n \prod_{i=1}^n 
\ell_i^{-(1+\g_0)}.
\eeq  
The factor $A^n\,Z^{(g=0)}(A)$ is uncontroversial, with 
$A^n$ counting the number of ways $n$ boundary loops can be located
on a surface of area $A$. In addition, JM made the general 
ansatz $1/\ell^{(1+\a)}$, where $\ell$ is less than some fraction of $\sqrt{A}$.
To determine $\alpha$, they
calculated the genus-one partition function
from $\tZ_2(A,\ell,\ell)$
by gluing together the two boundaries to form a torus 
and integrating
over $\ell$ up to $\sqrt{A}$, 
\beq\label{2b}
Z^{(g=1)}(A) \sim \int^{\sqrt{A}} d\ell \; \ell \;\tZ_2(A,\ell,\ell),
\eeq
where the factor $\ell$ in the integrand takes into account the number of different gluings
of the two boundary loops.
Comparing to \rf{1a} and \rf{2a} one obtains $\a =\g_0$.

A number of interesting results were derived in \cite{jm} using the ansatz 
\rf{1b}. For example, it is straightforward to obtain the relation
$\g(g)=\g_0+g(2-\g_0)$ by employing \rf{1b} for a surface $A$  
with $g$ pairs of boundaries of equal loop length
$\ell_i$, $i=1,\ldots,g$, and thus $n=2g$. 
General genus-$g$ surfaces are created by gluing 
together the loops of each pair and integrating over the corresponding
length variable $\ell_i$, generalizing \rf{2b}. It is easy to see that each such 
operation will create a factor $A^{2-\g_0}$.  

As pointed out in \cite{jm}, the number of spherical surfaces of area $A$
with $n$ boundaries 
which cannot be deformed without increasing their length is strictly smaller
than the number of surfaces when the $n$ boundaries are arbitrary, for which
standard formulas exist\footnote{What could have alerted the authors 
of \cite{jm} to the fact that they were not quite on the right track is that
their formula for $n \geq 3$ and $c=0$ is {\it identical} to 
the standard expression (without any restrictions on the loops)
derived in \cite{ajm}, namely,
$$
Z_n(A,\ell_i) \sim Z^{(g=0)}(A)\; A^n \;\Big( \ell_1\cdots \ell_n\Big)^{-1/2} 
\; \exp \left(-\frac{(\ell_1+ \cdots+\ell_n)^2}{4A}\right).  
$$
This formula is valid also for $\ell_i > \sqrt{A}$ and shows  
that the cutoff in $\ell$ for ordinary boundaries is of order 
$\sqrt{A}$, as one would expect from the dimensionalities of $\ell$ and 
$A$.}. This happens because boundaries which serve as baby universe necks 
in the sense of JM 
are {\it special} curves. However, the consequences of this are even more drastic
than envisaged in \cite{jm}, and are directly related to why 
the ansatz \rf{1b} is not entirely correct.
The boundary loop in \rf{1b} is by definition a geodesic curve,
and it is well known that geodesics scale anomalously in the DT ensemble
of surfaces \cite{kawai0,gk,aw,ajw,kawai1}. 
In fact, the dimension of geodesic curves is volume$^{1/4}$ 
and not volume$^{1/2}$, as one might have expected na\"ively, 
implying that the Hausdorff dimension $d_h$ of surfaces in the  
DT ensemble is 4. This is reflected in the scaling behaviour of the
expectation values (w.r.t. the ensemble average)
\beq\label{3b}
\la R\ra_A \sim A^{1/d_h},~~\la A(r) \ra_A \sim r^{d_h}, ~~r\ll A^{1/d_h},
\;\; {\rm where}\; d_h=4\;\;\; (c=0),
\eeq
where $R$ is the linear extension of a surface of area $A$,
and $A(r)$ the area contained within a geodesic distance $r$ from a given point. 

The necks of length $\ell$ along which JM
cut surfaces into disconnected pieces are geodesic curves, which means
that their ensemble average is much shorter than the generic $\sqrt{A}$
used in \cite{jm} as upper limit in integrations like \rf{1b}. (Note 
that the main contribution to this integral comes precisely from the upper limit.) 
Instead, according to \rf{3b}, $A^{1/4}$ should be used as upper limit, and, 
more generally, $A^{1/d_h}$ if the Hausdorff dimension is 
$d_h$. An alternative ansatz, 
which will reproduce most of the results of \cite{jm}, is 
to replace \rf{1b} by 
\beq\label{4b}
\tZ_n'(A,\ell_i)\sim Z^{(g=0)}(A) \; A^n\prod_{i=1}^n 
\ell_i^{-(1+\g_0d_h/2)}  \;\;\; (c=0).
\eeq  
This formula is supposed to be valid for $\ell_i < A^{1/d_h}$, with the
understanding that any integrations over $\ell_i$ are to be performed 
from the minimal neck length up to $A^{1/d_h}$. For $\ell_i > A^{1/d_h}$
the function $\tZ_n'(A,\ell_i)$ is assumed to vanish fast.

\section{Testing the new ansatz}

As an application of the new ansatz \rf{4b}, let us reconsider the torus case, 
where relation \rf{2b} is now replaced by 
\beq\label{5b}
Z^{(g=1)}(A) \sim \int^{A^{1/4}} d\ell \; \ell \;\tZ'_2(A,\ell,\ell).
\eeq
We have performed a nontrivial quantitative test of the ansatz \rf{4b} 
by measuring the probability distribution $P_A(\ell)$ of the length $\ell$ of the
shortest non-contractible loop 
in the DT ensemble of triangulations of the torus with $A$ triangles.
According to \rf{4b}, the distribution should be 
\beq\label{7b}
P_A(\ell) = \frac{\ell \;\tZ'_2(A,\ell,\ell)}{Z^{(g=1)}(A)} \sim 
\frac{\ell}{A^{1/2}} F\Big(\frac{\ell}{A^{1/4}}\Big), 
\eeq   
where $F(0)=1$ and $F(x)$ goes rapidly to zero when $x >1$.

In practice, to find shortest non-contractible loops we need a method to 
determine whether a given loop is contractible or not. By a different method,
we first constructed two loops $\alpha_1$, $\alpha_2$ generating the fundamental 
group of the torus. 
Whether an arbitrary loop is contractible can then be established by measuring 
its intersection numbers with these two generators. It is contractible if and only 
if both intersection numbers vanish.
The shortest non-contractible loop based at any given vertex $v$ can
be found by performing a so-called breadth-first search along the edges starting at $v$. 
Encountering a vertex that has been visited before means that one has
found a loop in the triangulation. The first such loop one meets which is  
non-contractible will automatically have minimal length. ÊÊ
Since any non-contractible loop must intersect at least one of the $\alpha_i$, 
in order to find the overall shortest non-contractible loop it suffices to repeat the above 
procedure for all vertices contained in either of the generating 
loops $\alpha_i$.\footnote{To optimize performance one should choose generating 
loops which are relatively short. In line with the result presented in this paper, 
this means that their length is of the order $V^{1/d_h}$, implying an 
expected run-time of the full algorithm of the order $V^{1+1/d_h}=V^{1.25}$.} ÊÊÊÊ

\begin{figure}[t]
\centerline{\scalebox{1.1}{\rotatebox{0}{\includegraphics{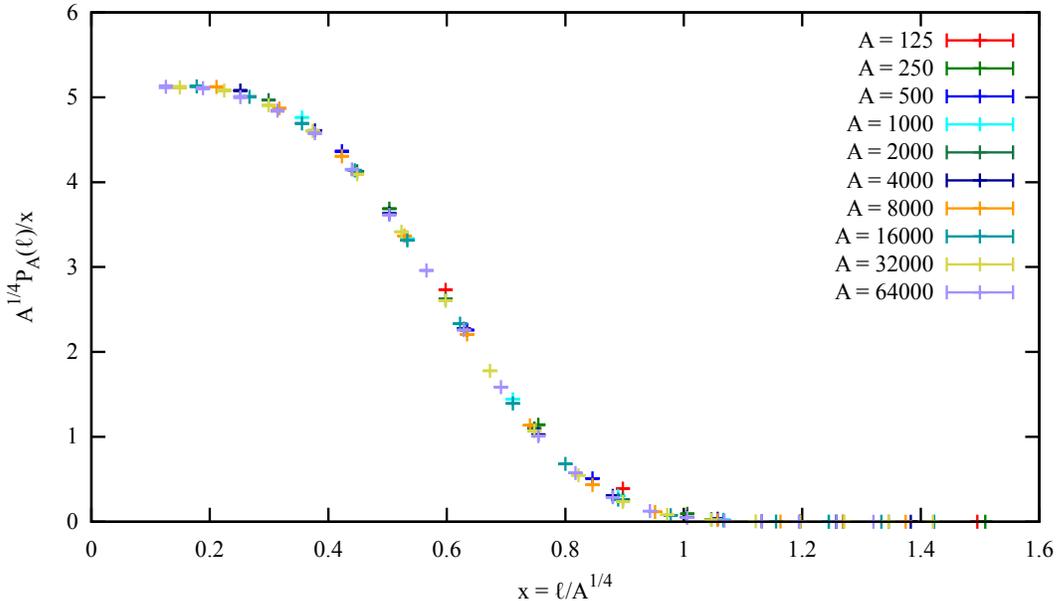}}}}
\caption[fig1]{Plot of a rescaled version $A^{1/4} P_A(\ell)/(\ell/A^{1/4})$
of the probability distribution of the
length of the shortest non-contractible loop on the DT ensemble of
triangulated tori, as function of the rescaled length $x=\ell/A^{1/4}$, 
with $\ell\in [2,21]$ and areas $A$ in the range between 125 and 64000.} 
\label{fig1}
\end{figure}

Fig.\ \ref{fig1} shows the superimposed measurements of the rescaled 
probability distribution
$A^{1/4} P_A(\ell)/(\ell/A^{1/4})$
for various values of $\ell$ and $A$.
The computer simulations were performed in the DT ensemble consisting of
triangulations dual to connected $\phi^3$-graphs with 
$A$ vertices and toroidal topology, including self-energy and tadpole graphs.
(To be precise, the simplest self-energy subgraphs consisting of two lines and 
tadpole subgraphs consisting of a single line were left
out to ensure that all vertices in the triangulation have
an order larger than or equal to three. This is not expected to affect the final
result in any way.)

If the ansatz \rf{4b} is 
correct we should observe a universal function of $x=\ell/A^{1/4}$
which goes to a constant for $x \to 0$. Inspection of Fig.\ \ref{fig1} reveals
that the expected finite-size scaling is indeed satisfied with good accuracy,
and we obtain a universal function $F(x)$.
By contrast, the computer simulations are in clear disagreement with 
the ansatz of \cite{jm}
which predicts a ($\ell$-independent) probability distribution 
\beq\label{8b}
P_A^{(JM)}(\ell) \sim \frac{1}{\sqrt{A}}~~~{\rm for}~~~\ell < \sqrt{A},
\eeq
as noted in \cite{thordur}. Our improved ansatz also resolves a discrepancy
found in \cite{thordur}, namely, that \rf{8b} leads to an average loop length  
$\la \ell \ra \sim A^{1/2}$, while one would expect $\la \ell \ra \sim A^{1/4}$
because of the anomalous scaling dimension of geodesic distance. It is exactly 
the latter relation which follows from our suggested probability distribution \rf{7b}, 
based on the ansatz \rf{4b}.

\section{2d gravity coupled to conformal matter}\label{sec3}

The arguments presented above for $c\! =\! 0$ need to be refined
for the general case $c\!\leq\! 1$, when matter is coupled to 2d gravity. 
The idea put forward in \cite{jm} of how to tackle this situation is to integrate
out all matter degrees of freedom and work with the resulting partition function
(which {\it looks} purely geometric, like the one for $c\! =\! 0$), and then simply repeat
the cutting-and-gluing arguments for baby universes used in the matter-free case. 
Although this is a well-defined procedure for a single closed surface of area $A$, and will lead 
to a partition function $Z^{(g)}(A)$, this is no longer true in the
presence of boundaries where, to start with, boundary conditions for the matter 
must be specified. Examples are so-called ``free boundary conditions"
(where the matter configurations on the boundary are integrated over), or other
specific
prescriptions like Dirichlet boundary conditions for the case of a scalar field, say.
No matter what choice one makes, additional weights associated with the matter
on the boundary will arise, which are simply not taken into account in the
treatment of \cite{jm}, which crucially features ``baby universe surgery" along such
boundaries. 

Let us illustrate this point by the case of Ising spins coupled to DT, with spins
located at the centres of triangles. The Ising model has a 
critical point when defined on the DT ensemble, 
describing a $c=1/2$ conformal field theory coupled to 2d gravity \cite{kazakov}. 
If we consider a surface containing a neck and an associated baby universe,
there is clearly an energy associated with the spin configurations of neighbouring 
triangles across the loop forming the neck, which will appear as a weight in the 
statistical sum over spin configurations. There is no obvious way to recover this
energy contribution exactly from the separate partition functions of the
two disjoint surfaces after cutting open the initial surface along the neck.

A perhaps more pertinent question is when and whether such unaccounted
boundary contributions to the energy will matter. 
In the case of minbus the boundaries in question will have a minimal number of 
edges (one, two, or three, depending on the choice of DT ensemble), and
any energy contribution associated with them will become negligible in the
continuum limit. This explains why the relations derived by JM are robust
for the case of baby universes of minimal neck size, even when $c\not= 0$.
As already mentioned in passing, their relation \rf{4a} for the average minbu
number has been used to measure $\g_0$ also in the matter-coupled case, 
and agreement with the theoretical value for $\g_0$ was found.

The derivations in \cite{jm} building on the ansatz \rf{1b} for general, non-minimal
necks are much harder to justify. When the boundary lengths $\ell$ 
become of the order of the linear system size (it is precisely those large 
$\ell$-values which give important contributions to integrals like \rf{2b}, 
as noted earlier), the contributions from boundary energies become
potentially significant, but are very difficult to control. Nevertheless, let us 
for the sake of simplicity assume that we can  
ignore such boundary energies when decomposing surfaces
and that a suitable generalization of the ansatz \rf{4b} is valid,
and see where this leads us. 

For definiteness, consider the Ising model on the DT ensemble
as described above. As long as the coupling constant of the Ising model
stays away from its critical value, the geometric behaviour of the model 
will be identical to that of the pure-gravity case with $c=0$ and $d_h=4$,
and captured by our earlier ansatz \rf{4b}, together with the new insight
that the linear size of the surface is not given by the `na\"ive' $A^{1/2}$, but 
by $A^{1/4}$. By contrast, if we fine-tune the Ising coupling to 
its critical value, the interaction between geometry and spins 
will be such that the fractal structure of spacetime is altered, manifesting
itself in a larger Hausdorff dimension $d_h({\it Ising})\! >\! 4$. This also implies that
the effective linear size of the system is decreased 
to $A^{1/d_h({\it Ising})}$. It makes it tempting to conjecture that a straightforward
generalization of \rf{4b}, namely,
\beq\label{4bgen}
\tZ_n'(A,\ell_i)\sim Z^{(g=0)}(A) \; A^n\prod_{i=1}^n 
\ell_i^{-(1+\g_0(c)d_h(c)/2)}  \;\;\; ({\rm any}\ c\leq 1),
\eeq  
is a suitable ansatz in the matter-coupled case too. It accommodates the change
in Hausdorff dimension, as well as reproducing the correct partition function
$Z^{(g=1)}(A)$ for the Ising model on a torus.
The general formula for $d_h$ as a function of the central charge $c$ of 
the conformal field theory to be used in \rf{4bgen} is \cite{watabiki} 
\beq\label{1c}
d_h(c)= 2 \frac{\sqrt{25-c}+\sqrt{49-c}}{\sqrt{25-c}+\sqrt{1-c}},
\eeq 
together with the expression given in \rf{2a} for $\g_0(c)$.
The relation which generalizes \rf{7b} is 
\beq\label{2c}
P_A(\ell)\sim A^{-1/d_h(c)} \Big(\frac{\ell}{A^{1/d_h(c)}}\Big)^{|\g_0(c) |d_h(c)-1}
F\Big(\frac{\ell}{A^{1/d_h(c)}}\Big),
\eeq
where $F(0)\! =\!1$ and $F(x)$ falls off fast for $x\! >\! 1$
(and recall that $\gamma_0(c)\!\leq\! 0$). We conclude that $A^{1/d_h(c)} P_A(\ell)$
should be a universal function $x^\a F(x)$ of the rescaled length variable
$\ell/A^{1/d_h(c)}$, with $\a = |\g_0(c)| \d_h(c)-1$. 

One could test the prediction \rf{4bgen} numerically for the Ising model,
where $\a$ should be 0.40. A more clear-cut test may be topological 
2d gravity, corresponding to $c\! =\! -2$, for which $\g_0\! =\! -1$,
$d_h\! =\! 3.56$ and thus $\a\! =\! 2.56$. A special feature of this case is 
that triangulated surfaces of spherical topology 
can be constructed recursively with the correct 
weight, removing the need for Monte Carlo simulations \cite{kk}.
This was employed in \cite{jaetal} to check \rf{1c} numerically with 
great precision. If one can generalize the
recursive generation of triangulated surfaces to toroidal topology 
it may be possible to obtain a high-precision determination of $\a$ too.   

Lastly, there exists a formula for $d_h(c)$ different 
from \rf{1c}, namely
\beq\label{3c}
d_h(c) = -\frac{2}{\g_0 (c)},
\eeq
dating all the way back to one of the original papers
of quantum Liouville theory \cite{dhk}. 
It agrees with \rf{1c} for $c\!=\!0$, where $d_h\! =\! 4$. However, it 
gives $d_h\! \to\! 0$ for $c\!\to\! -\infty$ (where one would expect $d_h\! =\! 2$),
and $d_h\! \to\! \infty$ for $c\! \to\! 1$. It has the interesting 
feature that the exponent $\a$ introduced above equals 1, not 
only for $c\! =\! 0$, as we have already verified, but for {\it all} $c\!  <\! 1$.
By now it is believed that the $d_h(c)$ of eq.\ \rf{1c}
does not reflect directly any geometric properties of the fluctuating
geometry, but rather critical aspects of the matter fields 
coupled to 2d gravity. This was made explicit in \cite{aajk},
where it was shown that the spin clusters of the matter fields 
had exactly the dimension \rf{3c} for both $c\! =\! -2$ and $c\! =\! 1/2$,
while measurements of the geometric $d_h$ clearly differed from \rf{3c}.
Measuring the exponent $\a$ for $c\! =\! -2$ and $c\! =\! 1/2$ as mentioned
above would give us a new method, compared to \cite{aajk}, to decide whether \rf{1c} or 
\rf{3c} represents the geometric Hausdorff dimension of 2d quantum gravity
coupled to matter.

\vspace{.5cm}      
\noindent {\bf Acknowledgements.} JA would like to thank the Institute of 
Theoretical Physics and
the Department of Physics and Astronomy at Utrecht University for hospitality 
and financial support.
TB and RL acknowledge support by the Netherlands Organisation for Scientific 
Research (NWO) under their VICI program.

\end{document}